\documentclass[a4,aps,prl,amsmath,amssymb,floatfix,twocolumn,superscriptaddress,showkeys]{revtex4-1}
\usepackage{graphicx}
\usepackage{hyperref}
\usepackage{dcolumn}
\usepackage{bm}
\usepackage{xcolor}
\hypersetup{
    colorlinks,
    linkcolor={red!80!black},
    citecolor={blue!50!black},
    urlcolor={blue!80!black}
}

\usepackage{epstopdf}
\usepackage{times}
\usepackage{mathptmx}

\begin{document}
\title{Quantifying long-term evolution of intra-urban spatial interactions}

\author{Lijun Sun}

\affiliation{Future Cities Laboratory, Singapore-ETH Centre for Global Environmental Sustainability (SEC), 138602, Singapore}
\affiliation{Department of Civil \& Environmental Engineering, National University of Singapore, 117576, Singapore}

\author{Jian Gang Jin}

\affiliation{School of Naval Architecture, Ocean and Civil Engineering, Shanghai Jiao Tong University, Shanghai, 200240, China}

 \author{Kay W. Axhausen}
\affiliation{Future Cities Laboratory, Singapore-ETH Centre for Global Environmental Sustainability (SEC), 138602, Singapore}
\affiliation{Institute for Transport Planning and Systems (IVT), Swiss Federal Institute of Technology, Z\"{u}rich, 8093, Switzerland}

\author{Der-Horng Lee}

\affiliation{Department of Civil \& Environmental Engineering, National University of Singapore, 117576, Singapore}

\author{Manuel Cebrian}

\affiliation{National Information and Communications Technology Australia, University of Melbourne, Victoria 3010, Australia}

\begin{abstract}
Understanding the long-term impact that changes in a city's transportation infrastructure have on its spatial interactions remains a challenge. The difficulty arises from the fact that the real impact may not be revealed in static or aggregated mobility measures, as these are remarkably robust to perturbations. More generally, the lack of longitudinal, cross-sectional data demonstrating the evolution of spatial interactions at a meaningful urban scale also hinders us from evaluating the sensitivity of movement indicators, limiting our capacity to understand the evolution of urban mobility in depth. Using very large mobility records distributed over three years we quantify the impact of the completion of a metro line extension: the circle line (CCL) in Singapore. We find that the commonly used movement indicators are almost identical before and after the project was completed. However, in comparing the temporal community structure across years, we do observe significant differences in the spatial reorganization of the affected geographical areas. The completion of CCL enables travelers to re-identify their desired destinations collectively with lower transport cost, making the community structure more consistent. These changes in locality are dynamic, and characterized over short time-scales, offering us a different approach to identify and analyze the long-term impact of new infrastructures on cities and their evolution dynamics.
\end{abstract}


\maketitle

\section{Introduction}

Cities, as the core of modern society, are playing increasingly important roles through global urbanization, providing people with housing, transportation, communication and functional institutions for various social activities. Enabled by the transportation infrastructure layer, diverse social interactions among various entities shape a city's interaction layers, creating social economic outputs, which further spur the growth of the cities themselves \cite{Bettencourt2007,Batty2008,Bettencourt2013,Pan2013}. Created by individuals' trips for work, school, shopping and other social activities, intra-urban movement is a crucial part of these spatial interactions. Intra-urban movements exhibit strong spatial and temporal patterns, which play an important role in urban planning and traffic forecasting \cite{Roth2011,Schlich2003,Axhausen2002}. Taking travel demand as an example, previous studies have focused on developing models to estimate current and predict future interaction intensity based on social and infrastructural input for planning purposes \cite{Zipf1946,Erlander1990,Stouffer1940,Wilson1969,Simini2012,Yan2014}. Although extensive efforts have been made using travel diary data  collected from household surveys and interviews \cite{Schlich2003,Axhausen2002}, studies on individual/collective movement still suffers from their small sampling shares, high cost, infrequent periodicity and limited accuracy. As a result, despite the observations telling us that near things are more related than distant things' geographically \cite{Tobler1970}, studies trying to integrate the infrastructure and interaction layers have remained limited due to the lack of detailed longitudinal data measuring change at high spatial and temporal resolutions.

The emerging individual-based large-scale data sets have allowed us to trace our daily behavior pattern in detail, shifting our understanding on individual mobility from random to predictable \cite{Gonzalez2008,Song2010,deMontjoye2013}. The smart card data also shows pronounced advantages in depicting the structure of collective encounter networks given the large share of public transit (capturing more than 63\% of total mobility in Singapore) \cite{Sun2013}. Besides revealing mobility regularity, these data sets also help us further understand human mobility-induced spatial interactions, which are crucial to various urban diffusion processes such as epidemic spreading, knowledge-spillover and social contagions \cite{Christakis2013,Balcan2009,Sun2014}. Thus, recent studies demonstrate an increasing use of network theory to model diverse types of interactions (from transportation to digital communication) on multiple scales (state, country and global) \cite{Barthelemy2011}, documenting the importance of such interactions among spatial units in shaping network typologies \cite{Barrat2004,Guimera2005,Ratti2010,Expert2011,DeMontis2013}. This paper analyses intra-urban movements using three one-week transit use data, including both bus and metro systems, from three years (April 11 to April 15, 2011; March 19 to March 23, 2012; and April 8 to April 12, 2013) in Singapore. We place special emphasis on the study of a key transportation infrastructure---the extended Circle Line (CCL; Mass Rapid Transit service), which provides us an ideal natural observation to investigate how such large infrastructures influence human mobility, and how the resulting spatial interactions shape the structural evolution of a city. The first stage of the CCL (the Eastern half; Fig. 4D, colored orange) has been in operation since 2010. After the competition of the second Western half, the CCL has been fully operational (35.7 km in total; Fig. 4E) since October, 2011. This metro line cost about 10 billion Singapore Dollars (roughly 8 billion US Dollars), carrying about half a million passengers daily today (0.37 million/day in stage 1 and 0.53 million/day after opening stage 2). It seems reasonable that such a large infrastructure project should affect local/global mobility patterns and city structure, by first influencing individual travel patterns, and then by prompting second-order effects such as new businesses and real estate. Indeed, such effects can be measured by looking at the usage patterns of the extension and the geographical space where it occurs \cite{Sung2014}. However, these approaches suffer notable limitations when analyzing city-wide movements. First, we miss how this local change affects global/city-scale patterns: we cannot tell whether the new service alters mobility patterns by merely looking at local travel patterns. Second, there might be changes of mobility patterns enabled by the new infrastructure we are not aware of yet. Taking advantage of emerging network analysis, this paper tries to present some steps to overcome this knowledge gap.

\begin{figure*}[t]
\centering
\label{Fig:fig1}
\includegraphics[scale=0.95]{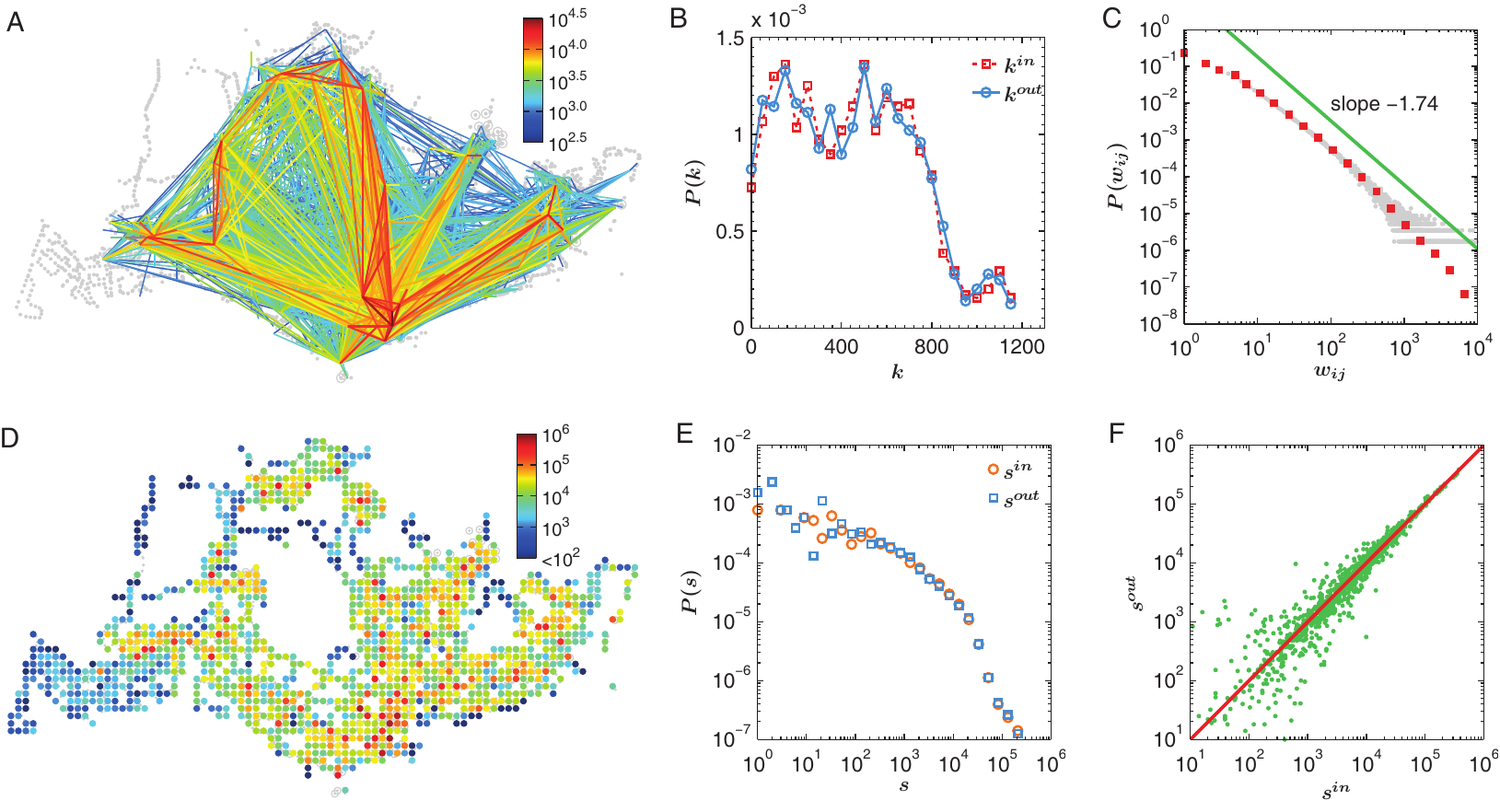}
\caption{Structure of intra-urban public transport movements in Singapore. (A) Aggregated spatial interaction network across weekdays (from March 19th to 23rd, 2012). For simplicity, we only show the top 1\% of edges with highest total interaction intensity ${{w}_{ij}}+{{w}_{ji}}$ in an undirected manner. The gray markers show the spatial locations of transit infrastructures, including both bus stops (dots) and metro stations (circles). (B) In/out-degree distribution $P\left( {{k}^{in}} \right)$ and $P\left( {{k}^{out}} \right)$ measured on the aggregated network. (C) Probability density function $P\left( {{w}_{ij}} \right)$ of interaction intensity across all OD pairs. Gray dots show the original histogram and red squares correspond to a log-binned histogram. As a guide, green line shows a power law with an exponent $\beta =1.74$. (D) The spatial distribution of total strength ${{s}_{i}}=s_{i}^{in}+s_{i}^{out}$ of each cell. (E) Probability density function $P\left( {{s}^{in}} \right)$ of in-strength and $P\left( {{s}^{out}} \right)$ of out-strength. Both of them exhibit heavy-tailed properties. (F) Symmetrical plot of ${{s}^{in}}$ and ${{s}^{out}}$. Most dots are scattered around the red line, suggesting the homogeneous inflow and outflow spatially.}
\end{figure*}

\section{Results}

To sketch the geographical structure of intra-urban movement from collective transit use, we divided the island city-state Singapore into zones of $500m\times 500m$, indexing all transit journeys with their origin-destination (OD) pairs (see SI Text for defining spatial zones and transit journey). By aggregating all transit journeys across weekdays using OD indexes in 2012, we obtained a directed and weighted spatial interaction network, which demonstrates symmetrical in/out-degree distributions (see Fig.~1A and Fig.~1B). This network---with spatial zones as its vertices and the time-resolved commuting flows as its edges---is the base of our analyses. To explore its statistical properties, we first measure interaction intensity ${{w}_{ij}}$ as total passenger flow traveling from zone $i$ to $j$, finding that when $w_{ij}\ge 10$, the tail of distribution $P\left( {{w}_{ij}} \right)$ is well characterized by a power-law $P\left( {{w}_{ij}} \right)\sim w_{ij}^{-\beta }$ with an exponent $\beta = 1.744\pm 0.002$ (Fig.~1C) using statistical tools provided in Ref. \cite{Alstott2014}. This indicates that intra-urban movement displays a strong heterogeneity: most ODs have small flows, but a few ODs involve massive demands. Note that a similar scaling is reported for London Underground \cite{Roth2011}, suggesting that such scaling of interaction intensity might be a fundamental property of urban spatial interactions. To quantify the importance of individual zones in shaping the aggregate interaction network, we measured in/out-strength of vertex $i$ as its total inflows and outflows ($s_{i}^{in}=\sum\nolimits_{i=1}^{N}{{{w}_{ji}}}$ and $s_{i}^{out}=\sum\nolimits_{j=1}^{N}{{{w}_{ij}}}$) on weekdays, respectively. As Fig. 1D shows, the spatial distribution of zone strength also exhibits a strong heterogeneity across the city. Although it is difficult to identify a deterministic function to characterize zone strength distributions $P\left( {{s}^{in}} \right)$ and $P\left( {{s}^{out}} \right)$, we still find the same heavy tail characterizing both of them, and observe an intrinsic balance across all zones (see Fig. 1E and Fig. 1F). Different from a previous study about the dependency of vertex strength $s$ on vertex degree $k$ in the worldwide airport network \cite{Barrat2004}, we found an exponential increase $s\left( k \right) \sim \exp \left( \lambda k \right)$ instead of a power-law $s\left( k \right) \sim {{k}^{\beta }}$ (with $\beta >1$), suggesting that the network demonstrates a much faster increase of $s$ than $k$ (see SI Fig~S1). In fact, considering the strong heterogeneity of ${{w}_{ij}}$, degree $k$, which merely counts the presence of edges and exhibits a saturation process with time, is not appropriate to capture the time-varying architecture and the backbone of this interaction network (SI Fig.~S2 and~S3). To check whether these properties hold over time, we applied these analyses for year 2011 and 2013 as well. Despite the completion of the CCL during the study period, we found indistinguishable aggregate properties for the three years.

\begin{figure*}[th]
\centering
\includegraphics[scale=0.95]{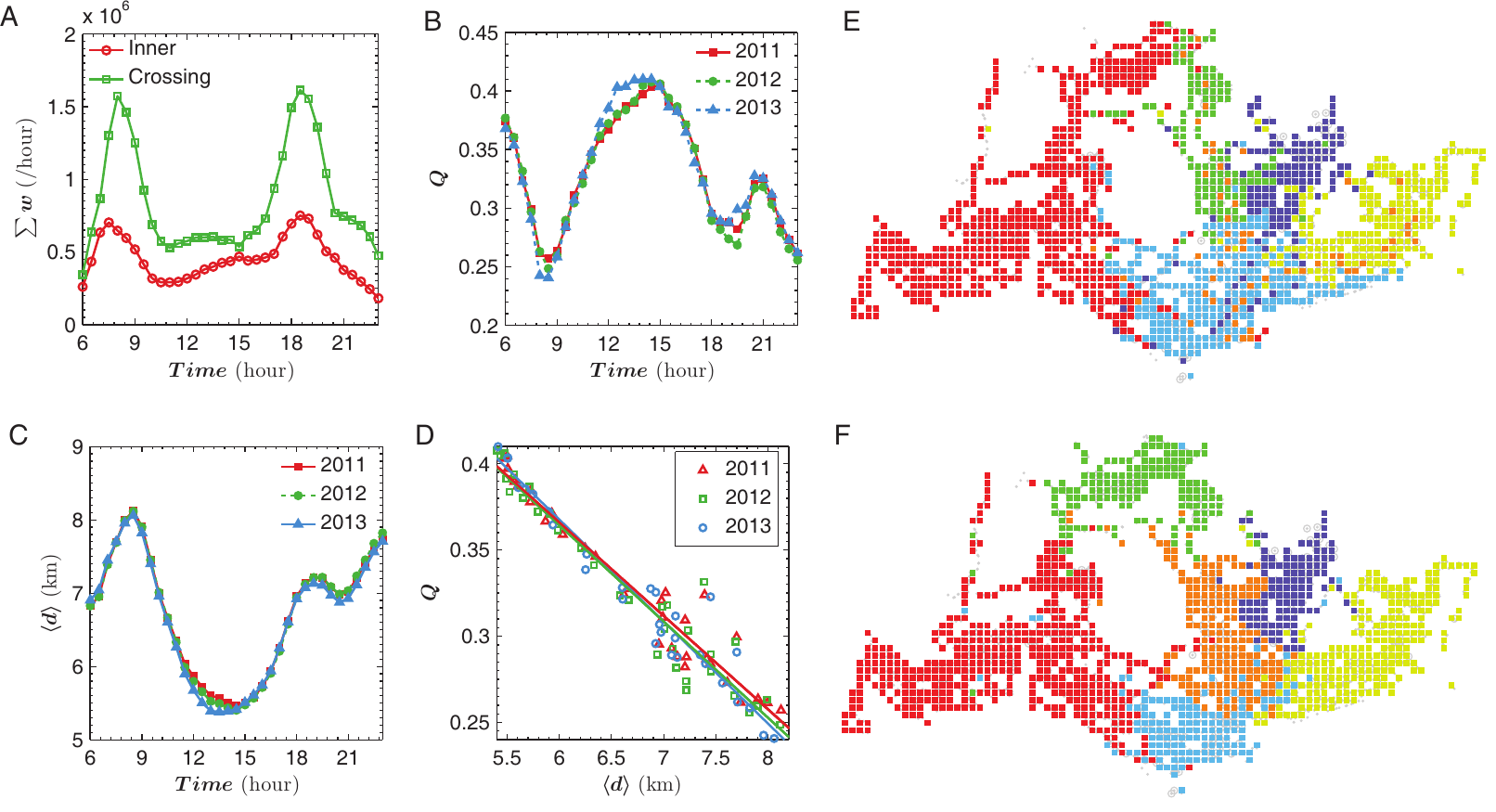}
\caption{Structure and temporal variation of intra-urban spatial interactions. (A) Temporal variation of total movements within communities (red-circled line) and across communities (green-squared line) in the year 2012. The summation of these two parts is the total transit flows. (B) Temporal change of modularity $Q$ in three years. Modularity $Q$ is characterized by a wave with pronounced troughs (at 8:30a.m. and 7:30p.m.) and peaks (at 2:30p.m. and 9p.m.). (C) Temporal variation of average trip distance $\left\langle d \right\rangle $ in three years. (D) The scatter plot of temporal modularity $Q$ and average trip distance $\left\langle d \right\rangle $. The solid lines represents linear regressions of $Q\sim \alpha \left\langle d \right\rangle $, with ${{\alpha }_{11}}=-0.69\pm 0.03$, ${{\alpha }_{12}}=-0.70\pm 0.04$ and ${{\alpha }_{13}}=-0.72\pm 0.03$ (the subscripts represent years), respectively. (E) and (F) Community structures identified for 8:30a.m. and 2:30p.m. in 2012.
}
\label{Fig:fig3}
\end{figure*}

A key property in understanding the dynamics of a spatial network is its community structure, defined as vertex partitions which have more connections within themselves than between each other. The importance of community lies in revealing the intermediate scales of network organization and identifying hidden structure in network theory. However, in practice the detection of communities is a difficult task to apply. The state-of-the-art approach to find these partitions is to maximize modularity $Q$ \cite{Newman2004}:
\begin{equation}
  Q=\frac{1}{W}\sum\limits_{ij}{\left( {{w}_{ij}}-{{w}_{ij}'} \right)\delta \left( {{c}_{i}},{{c}_{j}} \right)},
\end{equation}
where $W=\sum\nolimits_{i}{s_{i}^{in}}=\sum\nolimits_{j}{s_{j}^{out}}$ is the total network traffic, ${{{w}_{ij}'}}$ is the expected interaction intensity estimated from a null model and $\delta $ is an indicator function: $\delta \left( {{c}_{i}},{{c}_{j}} \right)=1$ if zone $i$ and $j$ belong to the same partition and $\delta \left( {{c}_{i}},{{c}_{j}} \right)=0$ otherwise. The approach has been successfully applied on various spatial interaction networks in different resolution, such as identifying effective administrative borders and finding hidden structures of countries and cities \cite{Guimera2005,Ratti2010,Expert2011,DeMontis2013}. In practice, an appropriate null model is crucial to get a meaningful expectation ${{{w}_{ij}'}}$ to reveal corresponding network structural attributes. Without accounting for spatial attributes, we adopted the default null model for defining modularity, i.e. ${{{w}_{ij}'}}=s_{i}^{out}s_{j}^{in}/W$ (see SI Fig.~S2). Note that one may replace the null model for special considerations; for example, ${{w}_{ij}'}$ is determined using a gravity model to exclude the dependency on distance in Ref. \cite{Expert2011}.

To find partitions that maximize modularity $Q$, we applied the well-established Louvain method on the aggregated network across weekdays \cite{Blondel2008}. Although the detection process employs only interaction intensity (without using any geographical information), we still observe a clear spatial consistency from 2011 to 2013 (see SI Fig.~S4), suggesting that collective movements are remarkably constant across years. Yet, we may miss the temporal evolution of these community structures by merely analyzing the aggregated network. To explore this evolution over the day, we grouped all transit journeys according to their transaction times (see SI Text for details). Based on these temporally grouped journeys, we created a series of sub-networks and applied the community detection processes on each of them. In Fig. 2A, we summed intra community and inter community flows and measured their temporal variations using 2012 data. Although community structure is well-defined spatially and intra interactions are much stronger than inter interactions ($\left\langle {{w}_{in}} \right\rangle >\left\langle {{w}_{out}} \right\rangle $), the total number of journeys crossing community is still higher than that of intra community trips ($\sum{{{w}_{in}}}<\sum{{{w}_{out}}}$), suggesting that people are not confined to a spatial community but show wider destination choices. Likewise, we repeated this analysis for the other years for comparison, finding that temporal variations of modularity $Q$ are essentially indistinguishable across years (Fig. 2B and SI Fig.~S6-S8). This indicates that temporal intra-urban movements (or collective mobility) might be comparable as well. Given the clear geographical consistence of these communities, modularity $Q$ is actually a measure of spatial mobility patterns embedded in temporal activities. To explore the variability of collective mobility over time, we measured the average trip distance $\left\langle d \right\rangle $ for each sub-network. Not surprisingly, we also found high degree of similarity of $\left\langle d \right\rangle $ across years (Fig. 2C), indicating that temporal travel behaviors of population are essentially comparable over three years as well. Next, we compared $Q$ and $\left\langle d \right\rangle $ jointly to explore how collective movement shapes the time-varying community structure. If communities are well identified as spatial clusters, the longer people travel, the higher the chance that one jumps out of a community. However, the relation is unclear without the assumption. For example, taking only structured metro trips will give to high modularity and long travel distance, whereas short and random local trips will lead to low modularity and short distance. We find that they are linked by a universal and substantial negative correlation $Q \sim \alpha \left\langle d \right\rangle $ (see Fig. 2D; with ${{\rho }_{2011}}=-0.9718$ [$p<{{10}^{-4}}$], ${{\rho }_{2012}}=-0.9623$ [$p<{{10}^{-4}}$] and ${{\rho }_{2013}}=-0.9763$ [$p<{{10}^{-4}}$]); and once again, we observed similar structural patterns across years, independently of the completion of the CCL. Therefore, using time-varying travel displacement as an indicator, we confirmed that temporal variation of collective movement plays a crucial role in expressing the dynamic community structures.

\begin{figure}[th]
\centering
\includegraphics[scale=0.95]{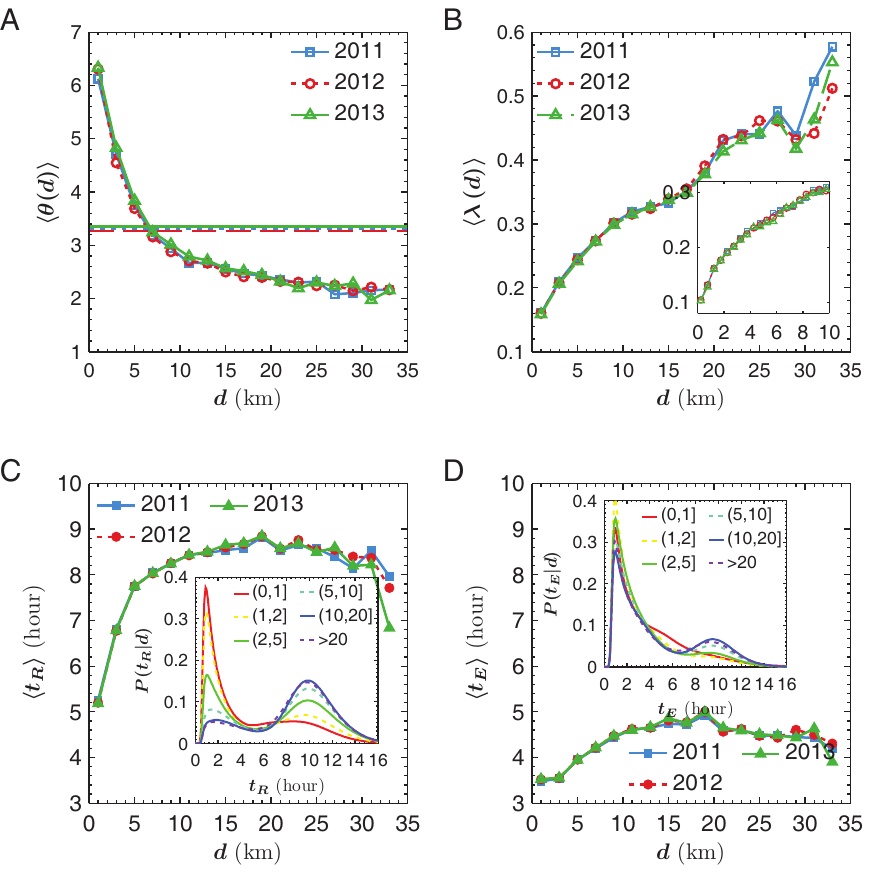}
\caption{The impact of distance on spatial interaction patterns. (A) The average diversity index ${{\theta }_{ij}}$ with given displacement ${{d}_{ij}}$. The dashed line demonstrations a null model assuming that $\theta $ is independent of $d$. (B) The average correlation $\left\langle {{\lambda }_{ij}} \right\rangle $ with ${{d}_{ij}}$. The inset illustrates the change of $\left\langle \lambda  \right\rangle $ when $0<d\le 10km$. (C) Stay duration ${{t}_{R}}\left( d \right)$ in between round trips $\left( i,j \right)$ and $\left( j,i \right)$, averaged over ${{d}_{ij}}$ for both years. The inset shows the distribution $P\left( {{t}_{R}}|d \right)$ for groups characterized by different $d$ using kernel smoothing. (D) The same plot as in panel (C), however, for the stay duration ${{t}_{E}}\left( d \right)$ in between trip chain. The inset shows the corresponding conditional distribution $P\left( {{t}_{E}}|d \right)$.
}
\label{Fig:fig3}
\end{figure}

To explore how distance affects systematic/occasional travel behavior, we quantified the variation of two distance-related diversity measures. Without considering individual identity, we first quantify the degree of heterogeneity of bi-directional flows by measuring a diversity index of OD $\left( i,j \right)$ as ${{\theta }_{ij}}=\max \left\{ {{w}_{ij}},{{w}_{ji}} \right\}/\min \left\{ {{w}_{ij}},{{w}_{ji}} \right\}$ (for OD pairs with ${{w}_{ij}},{{w}_{ji}}>0$). If the collective attribute ${{\theta }_{ij}}$ is independent of distance, we expect to find ${{\theta }_{ij}}$ being characterized by a determined distribution free from ${{d}_{ij}}$. However, on the contrary, we do find a significant and consistent reduction of $\left\langle {{\theta }_{ij}} \right\rangle $ with ${{d}_{ij}}$, indicating that bi-directional transit flows at a collective level become more and more balanced with distance. Still, this is not sufficient to show homogeneity at the individual level. Taking advantage of the anonymized card ID, we further investigated the similarity of users traveling between $i$ and $j$ for each day using the Jaccard index ${{\lambda }_{ij}}=\left| {{W}_{ij}}\cap {{W}_{ji}} \right|/\left| {{W}_{ij}}\cup {{W}_{ji}} \right|$, where ${{W}_{ij}}$ represents the set of individuals traveling from $i$ to $j$; thus, ${{\lambda }_{ij}}$ is close to one if all individuals travel symmetrically each day, and zero if no one returns to previously visited locations (i.e. ${{W}_{ij}}\cap {{W}_{ji}}=\varnothing $). After measuring ${{\lambda }_{ij}}$ for all OD pairs across weekdays, we show the dependence of $\left\langle {{\lambda }_{ij}} \right\rangle $ on ${{d}_{ij}}$ in Fig. 3B, finding another consistent (increasing) trend across three years. Therefore, despite previous observations characterizing exploration/preferential return behavior \cite{Song2010_2}, we found that shorter travel distances are associated with higher exploration; and correspondingly, previously visited locations are more preferred for longer distance journeys. Moreover, these properties are also stable across years, independent of the new metro line.

Large-scale human mobility patterns have been described by three indicators: trip distance distribution $P\left( d \right)$, temporal variation of radius of gyration ${{r}_{g}}\left( t \right)$ and number of visited locations $S\left( t \right)$. However, the duration of stay at one location, as another important attribute in understanding why people move rather than why people stay, is seldom considered in the literature. To investigate recurrence and periodicity of travel behavior and the patterns of stay in terms of both exploration/preferential return behaviors, we classified transit usage based on the pattern \cite{Schneider2013}: round journeys (with two trips $i\to j$ and $j\to i$) and trip chains with two trips $i\to j$ and $j\to k$ where $i\ne k$), and measured the duration of stay at zone $j$ for each. In Fig. 3C and D, we show the change of average duration of stay for both round journeys ($\left\langle{{t}_{R}}\right\rangle$) and exploratory trip chains ($\left\langle{{t}_{E}}\right\rangle$) as a function of distance ${{d}_{ij}}$, finding that both $\left\langle {{t}_{R}} \right\rangle $ and $\left\langle {{t}_{E}} \right\rangle $ display an consistent increase with $d$ in the beginning and reach saturation after $d=10km$. In comparing them, we find that ${{t}_{R}}$ is significantly longer than ${{t}_{E}}$ ($p<{{10}^{-4}}$, Wilcoxon rank-sum test), suggesting that people tend to stay longer at the destination of round trips. To further distinguish ${{t}_{R}}$ from ${{t}_{E}}$, we group journeys with similar travel distance ${{d}_{ij}}$ and determine the distribution $P\left( {{t}_{R}}|d \right)$ of stay duration (by measuring the interval between journey $i\to j$ and journey $j\to i$) and $P\left( {{t}_{E}}|d \right)$ (by measuring the interval between two journeys $i\to j$ and $j\to k$) for each group. As the insets of Fig. 3C and D show, both $P\left( {{t}_{R}}|d \right)$ and $P\left( {{t}_{E}}|d \right)$ can be approximately characterized by a mixture distribution of a short stopover (secondary activity around 1 hour, such as shopping, eating and leisure activities) and a long primary activity (around 10 hours, such as work and school). We note that the key difference between ${{t}_{R}}$ and ${{t}_{E}}$ is their composition: the proportion of primary activity in return journeys is significantly higher than that of a trip chain. Thus, Fig. 3C and D also imply a strong correlation between trip purpose and travel distance, further confirming the role of distance in shaping spatial interaction structure. Taken together, we show that travel distance not only determines the balance of intra-urban movement and a traveler's exploration/preferential return behavior, but also with the type of journey. Not surprisingly, we once again observe a clear consistency of ${{t}_{R}}$ and ${{t}_{E}}$ across the three years.

\begin{figure*}[th]
\centering
\label{Fig:fig4}
\includegraphics[scale=0.95]{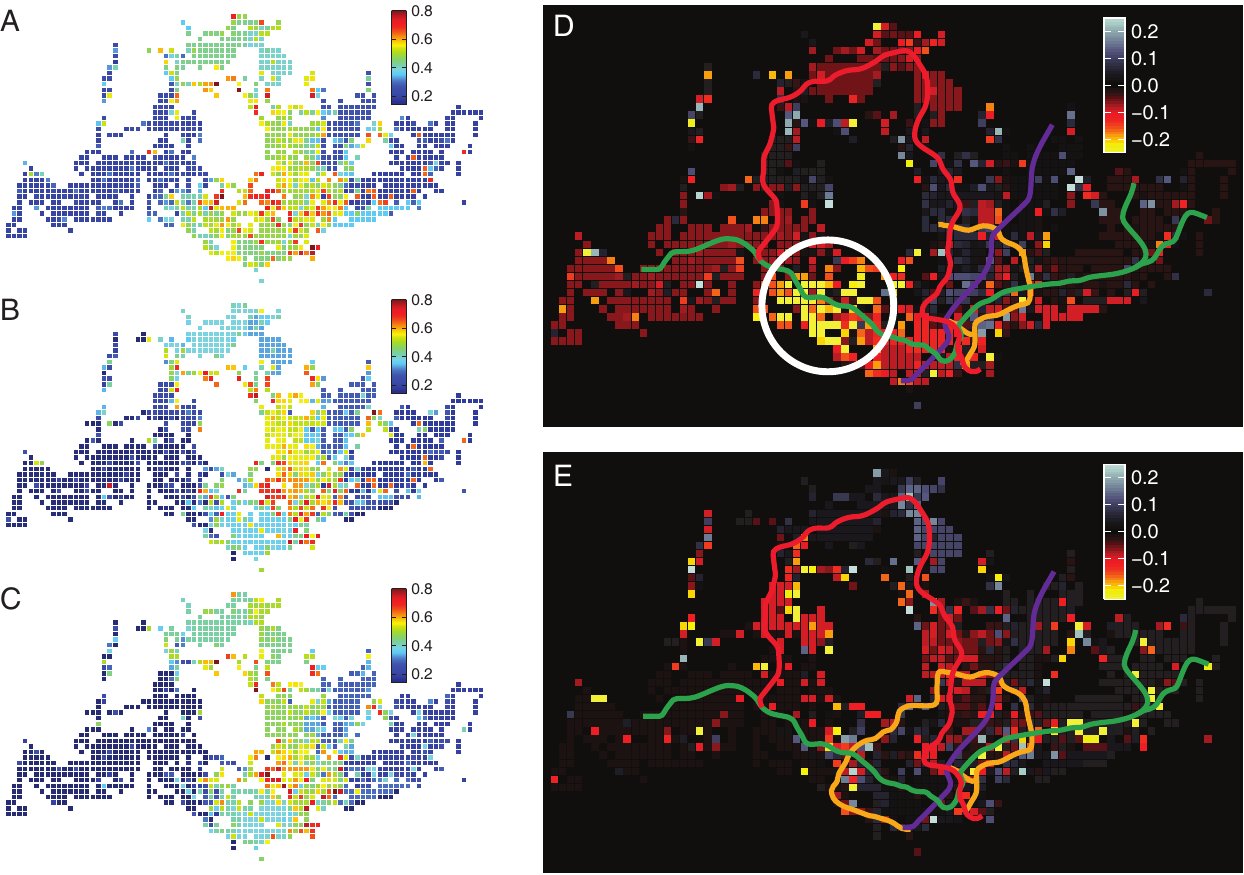}
\caption{Spatial distribution of mutability ${{\phi }_{i}}$ from 6:00 a.m. to 11 p.m: (A-C) ${{\phi }_{i,11}}$, ${{\phi }_{i,12}}$ and ${{\phi }_{i,13}}$ in the year 2011, 2012 and 2013, respectively. (D) The difference of mutability from 2011 to 2012 $\Delta {{\phi }^{1}}={{\phi }_{i,12}}-{{\phi }_{i,11}}$. (E) The difference of mutability from 2012 to 2013 $\Delta {{\phi }^{2}}={{\phi }_{i,13}}-{{\phi }_{i,12}}$. Given that the full completion of CCL was in October 2011, panel (A) shows the mutability of Stage 1 while both (B) and (C) show the mutability of Stage 2. Thus, panel (D) actually shows the change of mutability from Stage 1 to Stage 2, while both ${{\phi }_{i,12}}$ and ${{\phi }_{i,13}}$ are in Stage 2 for panel (E).
}
\end{figure*}

Pervious analyses has shown that intra-urban movement is consistent over the three years, exhibiting patterns independent of the key infrastructure project, suggesting that human mobility may display universal patterns invariant to the change of transportation infrastructures. However, as mentioned, with time-varying interactions during a day, the resulting structural communities should be changing simultaneously over time (such as the continuous changing of boarders and emergence of new communities). This effect offers us more insight into how spatial interactions shape the city. To quantify it, we define neighborhood variability ${{\gamma }_{i}}\left(t_1,t_2\right)$ of zone $i$ between the sub-networks at time ${{t}_{1}}$ and ${{t}_{2}}$ as:
\begin{equation}
\gamma_i\left(t_1,t_2\right)\equiv 1-\frac{\left|C_i\left(t_1\right)\cap C_i\left(t_2\right)\right|}{\left|C_i\left(t_1\right)\cup C_i\left(t_2\right)\right|},
\end{equation}
where ${{C}_{i}}\left( t \right)$ represents the community which zone $i$ belongs to at time $t$. Hence, ${{\gamma }_{i}}\left( {{t}_{1}},{{t}_{2}} \right)$ is close to one if the intersection contains only $i$ and zero if ${{C}_{i}}$ does not change from time ${{t}_{1}}$ to ${{t}_{2}}$. Using continuous observations during one day, we quantified the overall spatial evolution of zone $i$ by calculating mutability index ${{\phi }_{i}}$ as average neighborhood variability from ${{t}_{0}}$ till ${{t}_{max}}$ \cite{Palla2007}:
\begin{equation}
\phi_i\equiv
\frac{\sum_{t=t_0}^{t_{max}-1}\gamma_i\left(t,t+1\right)}{t_{max}-t_0},
\end{equation}
where ${{t}_{0}}$ is the time step when we start observing the temporal evolution and ${{t}_{max}}$ is the last time step. Thus, ${{\phi }_{i}}$ quantifies the overall evolution of community structure for each individual zone, characterizing the robustness/fragility of the spatial community to which zone $i$ belongs. Essentially, ${{\phi }_{i}}$ quantifies community transition rate of zone $i$ when other zones are also changing simultaneously. A high ${{\phi }_{i}}$ typically indicates that zone $i$ is attached to diverse communities over time and vice versa. In other words, the mutability index ${{\phi }_{i}}$ used here could also be interpreted as a measure to quantify the diversity of temporal community attachment. By setting ${{t}_{0}}=6$ and ${{t}_{\max }}=23$ (from 6a.m. to 11p.m. in 1 hour intervals, see SI Text for details), we determined the value of ${{\phi }_{i,y}}$ for each individual zone $i$ in year $y$. As shown in Fig. 4A, B and C, we find that mutability displays clear and comparable spatial patterns: one can easily distinguish the borders between regions with different mutability. Particularly, the central/southern area displays a higher mutability and the western/eastern parts are generally stable in three years.

To further compare mutability across years, we calculated $\Delta \phi $ (as $\Delta {{\phi }^{1}}={{\phi }_{i,12}}-{{\phi }_{i,11}}$ and $\Delta {{\phi }^{2}}={{\phi }_{i,13}}-{{\phi }_{i,12}}$) and map the results in Fig. 4D and E, respectively. Notably, although the temporal change of $Q$, $\left\langle d \right\rangle $ and other collective mobility indicators are essentially indistinguishable, we do observe a significant difference when using $\Delta {{\phi }^{1}}$ as an indicator, while no much difference is observed when measuring $\Delta {{\phi }^{2}}$. We think that the completion of the CCL appears as a main factor for this difference (see Fig. 4D). As mentioned, only the right half of the Circle Line was in operation in 2011 (Stage 1), while the full metro line has come into service since October, 2011 (Stage 2). Given the definition of $\phi_i$, the implication of $\Delta\phi_i>0$ are twofold. One one hand, for those zones have not changed their membership during a day --- such as most zones in the eastern/western community, the decrease of $\phi_i$ suggests that the community to which zone $i$ belongs becomes more consistent over time. On the other hand, for those zones changed their membership given time, a decreasing $\phi_i$ implies that zone $i$ changes less frequently and strengthened its dependency to the attached community. In fact, the completion of Stage 2 (the western half of the circle) enables travelers to find their desired destination choices collectively in a structured manner with lower cost, instead of making diverse choices individually. In this sense, the completion of Stage 2 of CCL may make zones in southern area (the white circle in Fig. 4D) more accessible to either the western community or the central community (as shown in Fig.~S6-S8). We next perform statistical test on $\Delta \phi_i$ for zones within the white circle (radius 4km, 124 zones nearby the extended CCL) and zones outside the circle (1170 zones). We find that $\Delta \phi _{i}^{1}$ in the nearby area are significantly lower than others ($p<{{10}^{-4}}$; left side Wilcoxon rank-sum test), while there is no clear evidence to show $\Delta \phi _{i}^{2}$ in the nearby area are different from others ($p=0.213$; Wilcoxon rank-sum test).

\section{Discussion}
Understanding spatial interactions is crucial to urban planning, traffic forecasting, and various mobility-related urban diffusion processes such as epidemic spreading and social contagions. More importantly, coupled with the transportation infrastructure layers of a city, social economic outputs are shaped by these interactions \cite{Fujita1999,Eagle2010}. Although the study on mobility has a long history, previous works were almost all based on modeling these interactions for planning purposes owing to a lack of longitudinal data detailed enough across both spatial and temporal scales. Yet, the evolution of urban structure with these temporal interactions is merely revealed: it remains a challenge to distinguish the natural variability in the city's mobility from large deviations, using either coarse-grained or short-time scales mobility data.

Taking advantage of population-scale smart card data sets spanning three years, we study the structure of Singapore's intra-urban interaction network and present how it is influenced by a key transportation infrastructure project (the CCL in this case). Despite that Singapore has been a dynamic, fast-changing city, we show that human mobility displays invariant aggregate patterns across the years, even when seeing a large infrastructure project. As a city evolves over the years, how can we distinguish large deviations in mobility from statistical fluctuations in a city's mobility patterns? Our study suggests that commonly used tools and statistics do not offer sufficient sensitivity to identify key changes of the city's mobility structure.

We present evidence for this by first examining the temporal community structure that emerges from collective travel behavior, and study its variation across years. Using modularity as an indicator, we find that the community structure varies consistently with the spatial-temporal characteristics of collective mobility, indicating that distance acts as a powerful constraint to keep universal mobility patterns in place over three year period, and therefore, does not allow us to discriminate the impact of the completion of the CCL. Moreover, we found that both modularity $Q$ and average journey distance $\left\langle d \right\rangle $ demonstrate clear and consistent temporal homogeneity, exhibiting remarkable robustness to the competition of the extended CCL as well. Taking stay-durations as another indicator quantifying human mobility, we showed that travel distance not only determines individual's exploration and preferential return behavior, but also associates with one's purpose of traveling. However, none of these indicators help us identify the global impact of CCL.

Notably, despite other structural and behavioral dynamic indicators being almost consistent and indistinguishable over the long-term, we do observe a significant difference of mutability. Mutability, which is defined as the average ratio of community members changed across time, emerges as a highly sensitive tool to understand the position of individual zones in the overall evolution and real-time evolving borders of community structure. In fact, it is sensitive to the differences in mobility caused by major transportation infrastructure change, showing the evolving borders in community structure, and with this the way people interact to shape, sustain or reform a city.

Our findings and analysis framework offers analytical tools to better sense the evolution of mobility patterns in cities, providing insights for urban planning, modeling and understanding the evolution of cities through the coupling of infrastructure and interaction networks. Given that ICT is being fast embedded in our daily-life, spatial-temporal digital traces helping us to follow cities will become available in various forms, overcoming the limits of field surveys and interviews. In the near future, much urban data which is generated in real-time will be become available for urban planning, improving the quality of life. Despite the privacy concerns, taking full advantage of such data in planning would surely help us further understand urban dynamics and make our cities smarter. Taken together, our study offers a quantitative and general strategy to understand the dynamic evolution in multiple temporal scales, and serve as a basis to further track and model such evolution \cite{Hopcroft2004}.

\section{Acknowledgments}
We thank M.Gonz\'{a}lez for discussions and comments on the manuscript. We thank Singapore's Land Transport Authority for providing the smart card data. This study was supported by National Research Foundation of Singapore, which is the funding authority of the Future Cities Laboratory, Singapore-ETH Centre. Manuel Cebrian is funded by the Australian Government as represented by the Department of Broadband, Communications and Digital Economy and the Australian Research Council through the ICT Centre of Excellence program.



\end{document}